\documentclass[sigconf]{acmart}
\AtBeginDocument{%
  }
    
\usepackage{CJKutf8}
\newcommand{\ja}[1]{\begin{CJK}{UTF8}{ipxm}#1\end{CJK}}

\copyrightyear{2026}
\acmYear{2026}
\setcopyright{cc}
\setcctype{by}
\acmConference[CHI EA '26]{Extended Abstracts of the 2026 CHI Conference on Human Factors in Computing Systems}{April 13--17, 2026}{Barcelona, Spain}
\acmBooktitle{Extended Abstracts of the 2026 CHI Conference on Human Factors in Computing Systems (CHI EA '26), April 13--17, 2026, Barcelona, Spain}
\acmDOI{10.1145/3772363.3798606}
\acmISBN{979-8-4007-2281-3/2026/04}

\begin{document}

\title{Shape vs. Context: Examining Human--AI Gaps in Ambiguous Japanese Character Recognition}



\author{Daichi Haraguchi}
\email{haraguchi_daichi_xa@cyberagent.co.jp}
\orcid{0000-0002-3109-9053}
\affiliation{%
  \institution{CyberAgent}
  \city{Tokyo}
  \country{Japan}
}


\begin{abstract}
High text recognition performance does not guarantee that Vision-Language Models (VLMs) share human-like decision patterns when resolving ambiguity. We investigate this behavioral gap by directly comparing humans and VLMs using continuously interpolated Japanese character shapes generated via a $\beta$-VAE. We estimate decision boundaries in a single-character recognition (shape-only task) and evaluate whether VLM responses align with human judgments under shape in context (i.e., embedding an ambiguous character near the human decision boundary in word-level context). We find that human and VLM decision boundaries differ in the shape-only task, and that shape in context can improve human alignment in some conditions. These results highlight qualitative behavioral differences, offering foundational insights toward human--VLM alignment benchmarking.
\end{abstract}


\begin{teaserfigure}
\centering
  \includegraphics[width=0.9\textwidth]{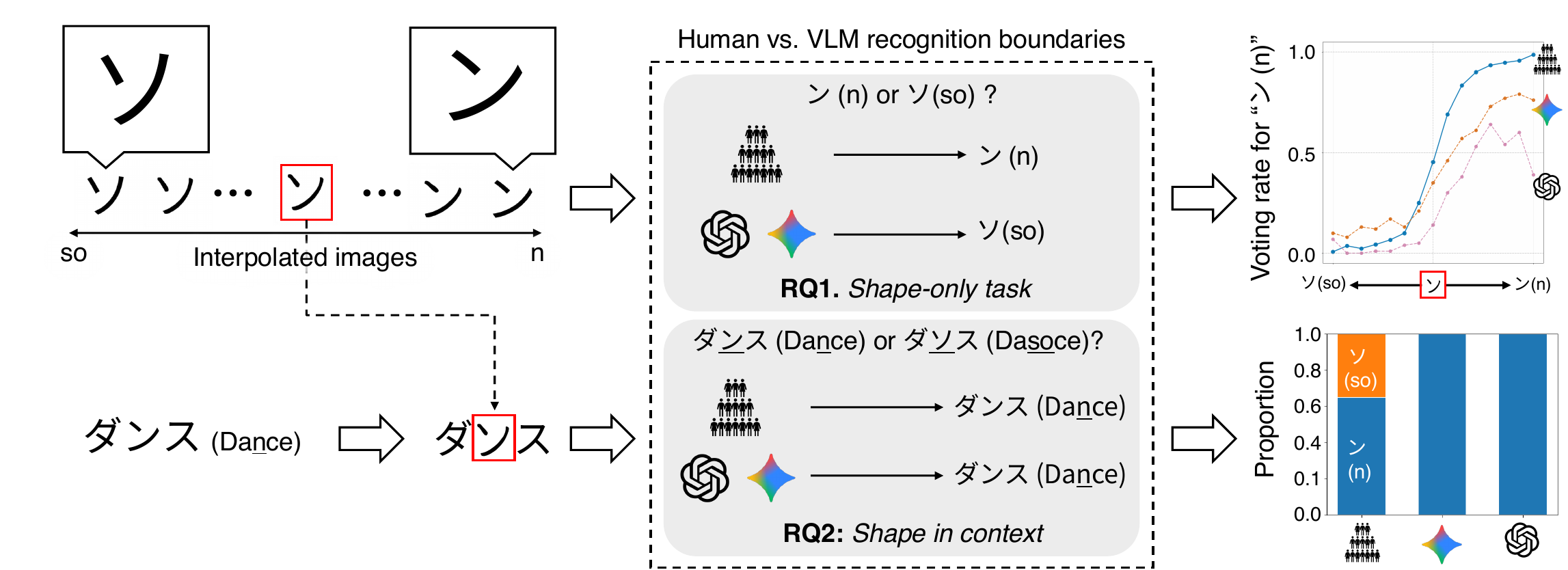}\\[-2mm]
  \caption{Probing Human--VLM alignment. Using interpolated images of the visually similar Japanese characters  `\ja{ソ}' (\textit{so}) and `\ja{ン}' (\textit{n}), we visualize the behavioral gap in two scenarios: RQ1 compares decision boundaries in a single character recognition (shape-only task), while RQ2 tests whether VLM responses align with human judgments in shape-in-context conditions.}
  \Description{teaser}
  \label{fig:teaser}
\end{teaserfigure}


\maketitle

\section{Introduction}
Large Vision-Language Models (VLMs), such as the GPT series~\cite{singh2025openai} and Gemini~\cite{team2023gemini}, can recognize text in images with high accuracy~\cite{liu2024ocrbench,fu2024ocrbench}. 
However, high recognition accuracy does not guarantee that these models make decisions in the same way humans do~\cite{firestone2020performance}. 
In particular, when visual evidence is uncertain, humans can flexibly use context to disambiguate input~\cite{cutler2024word,reicher1969perceptual,wheeler1970processes}, but it remains unclear whether VLMs exhibit similar flexibility or instead follow different, potentially biased decision patterns.
To characterize this behavioral gap, we compare human and VLM decision boundaries in a single-character recognition (shape-only task) and evaluate whether VLM responses align with human judgments in \textit{shape-in-context} conditions.
Understanding such differences is important for assessing how AI systems may behave under ambiguity in real-world settings, where interpretations can influence trust and downstream decisions.

As a controlled testbed for visual ambiguity, we focus on the Japanese characters `\ja{ソ}' (romanized as ``so,'' hereafter \textit{so}) and `\ja{ン}' (romanized as ``n,'' hereafter \textit{n}) (Fig.~\ref{fig:teaser}, top-left). 
This pair differs primarily along a single geometric dimension (stroke angle) while remaining nearly identical overall, making it well-suited for probing graded ambiguity and context-driven disambiguation. 
To move beyond static datasets, we use a $\beta$-VAE~\cite{higgins2017betavae} to generate a continuum of character images that smoothly interpolate between \textit{so} and \textit{n}, enabling fine-grained scans of decision boundaries for both humans and VLMs.

HCI research has extensively examined text legibility across typographic factors~\cite{palmen2023bold,wallace2022towards,cai2024cor} and viewing conditions~\cite{li2019impact,niklaus2023digital,rzayev2021reading,de2024caption}. 
This line of work is informed by cognitive models of how humans combine visual shape information (e.g., categorical perception~\cite{yasuhara1978category}) with contextual constraints during recognition (e.g., interactive activation~\cite{mcclelland1981interactive}). 
Against this background, we examine VLM behavior in character recognition by comparing VLM responses with human judgments under controlled ambiguity and context.

To this end, we address the following two Research Questions (RQs):

\textbf{RQ1:} How do the decision boundaries of VLMs differ from those of humans when processing isolated character shapes (shape-only)?

\textbf{RQ2:} In shape-in-context conditions, do VLM responses align with human judgments?

Our contribution goes beyond accuracy benchmarking by mapping decision boundaries in the shape-only task and testing human--VLM alignment in shape-in-context conditions. 
Although we focus on Japanese characters as a minimal test case, our results inform broader benchmark design for human--VLM alignment.

\section{Stimuli Generation via $\beta$-VAE}
\paragraph{Overview}To generate stimuli with continuously controllable visual ambiguity, we synthesized interpolated character images between \textit{so} and \textit{n} using a $\beta$-VAE~\cite{higgins2017betavae}. 
First, we trained the $\beta$-VAE on a large corpus of character images to learn a latent representation of character shapes. 
We then extracted latent representations of \textit{so} and \textit{n} and linearly interpolated between them in the latent space. 
By decoding these interpolated latent vectors, we generated a sequence of interpolated character images spanning 15 evenly spaced samples along the continuum between \textit{so} and \textit{n}.

\paragraph{Training Procedure and Stimuli Synthesis}
We prepared a dataset of 364 fonts from Google Fonts~\footnote{\url{https://github.com/google/fonts}} covering both serif and sans-serif styles across Japanese and Latin scripts~\footnote{In Japanese typography, these roughly correspond to Mincho (serif-like) and Gothic (sans-serif-like) styles.}.
Including diverse scripts allowed the model to learn a broad distribution of strokes. 
We trained a $\beta$-VAE (latent dim=32, $\beta=3.0$) on grayscale raster images ($256\times256$) using the Adam optimizer (learning rate=$10^{-3}$, batch=64) for 500 epochs, selecting the checkpoint with the lowest validation loss.

Using the trained $\beta$-VAE, we generated interpolated images between \textit{so} and \textit{n}. 
Latent representations of \textit{so} and \textit{n} were extracted using the encoder, denoted as $z_{\text{so}}$ and $z_{\text{n}}$. 
Interpolated images were obtained by decoding $z_{\text{inter}} = (1-\alpha)z_{\text{so}} + \alpha z_\text{n}$ with the decoder, where $\alpha \in [0, 1]$ is the interpolation parameter. 
We generated discrete stimuli by sampling $\alpha$ at 15 equally spaced intervals (including endpoints 0.0 and 1.0).

\paragraph{Construction of Contextual Word Images}
\begin{figure}[t]
    \centering
    \includegraphics[width=\linewidth]{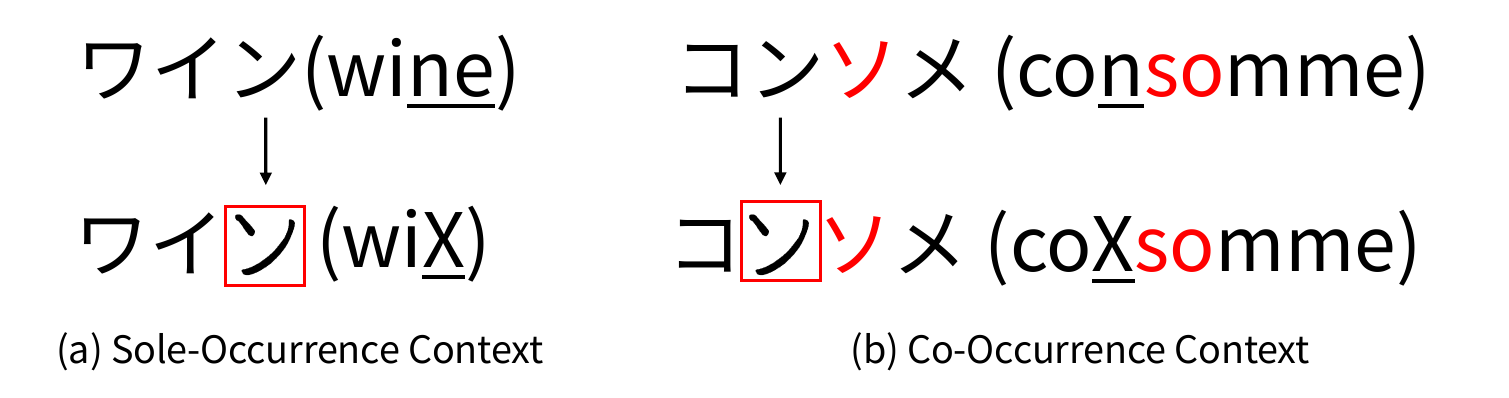} \\[-3mm]
    \caption{Construction of contextual word images.
Examples of sole-occurrence and co-occurrence contexts, where a single character is replaced with an ambiguous glyph (highlighted in red box) selected to yield approximately 50\% recognition by human participants in the shape-only task (RQ1).
In the co-occurrence context, additional \textit{so} or \textit{n} characters appear elsewhere in the word.}
    \label{fig:data}
\end{figure}
We constructed contextual word images by replacing a single character in each word with a visually ambiguous glyph (denoted as \textit{X}), rendered in Noto Sans JP (Fig.~\ref{fig:data}). 
The glyph \textit{X} was selected from the same latent interpolation series as in RQ1 experiment at the point of maximal ambiguity (approximately 50\% human recognition, at $\alpha = 0.429$), so that context could meaningfully influence decisions. 
At this point, however, the VLMs exhibited different shape-only priors: Gemini chose \textit{n} in about 40\% of repeated queries, whereas GPT never chose \textit{n} (0/10), indicating a strong bias toward \textit{so} in the RQ1 experiment.

We prepared 24 contextual conditions: 12 sole-occurrence~\footnote{Used word list: \ja{カワウソ, ピカソ, ワイン, ダンス, ソックス, ソファ, エスプレッソ, エピソード, カバン, ミカン, ライオン、ラーメン}.} and 12 co-occurrence~\footnote{Used word list:\ja{パソコン, コンソメ, ソフトバンク, インソール, ハンドソープ, マラソン}. Each base word was used twice (so-biased and n-biased), yielding 12 co-occurrence conditions.}. 
In sole-occurrence words, no other unambiguous instances of \textit{so} or \textit{n} appear outside \textit{X}, so constraints arise primarily at the lexical level. 
In co-occurrence words, additional \textit{so} and/or \textit{n} characters appear elsewhere in the word, providing within-word character cues. 
This design allows us to compare judgments when context is relatively weak versus when it is reinforced by co-occurring similar characters.

\section{Experimental Procedure for Humans and VLMs}
\subsection{User Study}
\paragraph{Shape-only task (RQ1)}
We recruited 30 participants. On each trial, participants viewed a single character image sampled from the \textit{so}--\textit{n} interpolation (including endpoints) and selected whether it was \textit{so} or \textit{n}. 
Each participant completed 150 trials (10 fonts $\times$ 15 $\alpha$ levels). 
To counterbalance order effects, we prepared three questionnaire forms with identical stimuli in different random orders, and participants were evenly assigned to one form.
Thus, we obtained 30 responses for each unique single-character stimulus.

\paragraph{Shape-in-context task (RQ2)}
We recruited approximately 390 participants. On each trial, participants viewed a word image in which one character was replaced with an ambiguous glyph \textit{X} (an interpolation between \textit{so} and \textit{n}). 
Rather than labeling $X$ directly, participants selected the intended reading of the whole word from a multiple-choice list, enabling comparison between human and VLM judgments under shape in context.

We used two context conditions. In the sole-occurrence condition, the word contained no other unambiguous instances of \textit{so} or \textit{n} besides  \textit{X}. 
In the co-occurrence condition, the word additionally included an unambiguous \textit{so} and/or \textit{n} elsewhere in the same word.
Answer options were the plausible whole-word readings for the stimulus (plus an ``Other'' option): 3 choices in the sole-occurrence condition and 5 choices in the co-occurrence condition.
For analysis, we aggregated responses at the ambiguous-character level rather than by whole-word choice: each selected reading was mapped to the implied identity of \textit{X} (i.e., \textit{so}, \textit{n}, or ``Other''). 
For example, for \textit{wiX} (based on \textit{wine}), choosing \textit{wine} was counted as an \textit{n} response, whereas choosing \textit{wiso} was counted as an \textit{so} response.

To support the independence assumption in our statistical tests, each participant evaluated at most one target word per condition (up to two target trials total), with unrelated filler trials interleaved.
We report sole-occurrence and co-occurrence results separately. Overall, each word--condition stimulus was evaluated by approximately 30 independent participants.

\paragraph{Ethics}
Each study consisted of a short (5–10 minute), anonymous online survey conducted via a Japanese crowdsourcing platform with monetary compensation.
No explicit language screening was applied. All study materials were presented in Japanese.
The task involved non-sensitive, forced-choice questions and did not include any deception, free-text responses, or collection of personally identifiable or detailed demographic information (e.g., age, gender, ethnicity). 
Participants were informed in advance that their responses may be used for academic and public dissemination. 
All responses were analyzed in aggregate, and the study design was intended to pose minimal risk to participants and minimize privacy concerns. In accordance with the affiliated institution’s guidelines, formal ethics review was not required for this study.


\subsection{VLM Experiments}
We evaluated VLMs under the same shape-only and shape-in-context conditions as in the human study. 
For each image, we prompted the model to choose from the same predefined answer options as humans (character identity in RQ1; whole-word reading in RQ2), and treated the selected option as the model response. 
For each stimulus, we issued 10 independent queries per model at temperature $1.0$ and aggregated responses for analysis. 
We report results for GPT-5.1 and Gemini-2.5-Flash.

\section{Analysis}
\subsection{Shape-only Character Recognition (RQ1)}\label{sec:rq1}
\begin{figure}[t]
    \centering
    \includegraphics[width=\linewidth]{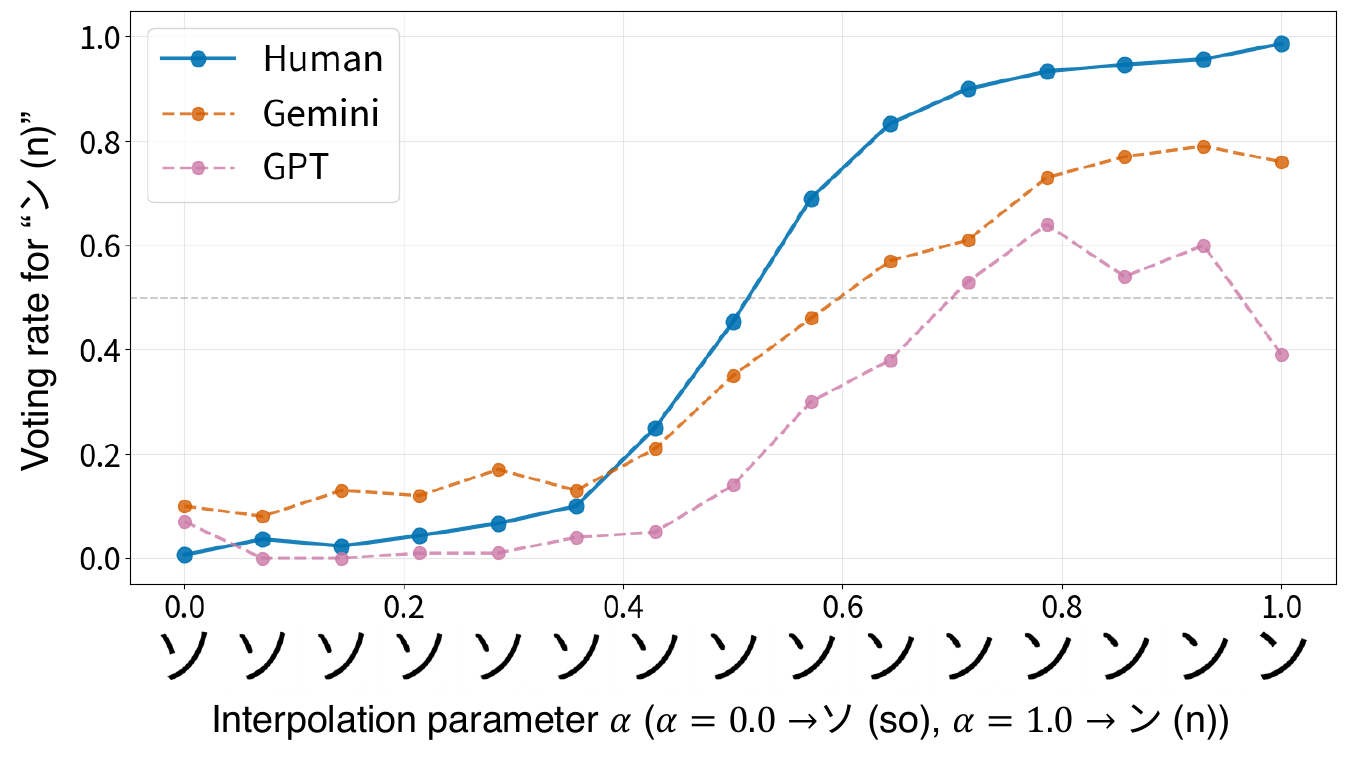}\\[-2mm]
    \caption{Single character recognition by each interpolation parameter (RQ1), aggregated across 10 fonts. Representative interpolated characters from one example font are shown along the x-axis for visual reference.}
    \label{fig:rq1}
\end{figure}

We analyzed shape-only responses with logistic mixed-effects models (random intercepts for participant and font) and tested the interpolation $\alpha \times$ VLM interaction via likelihood-ratio tests. The interaction was significant for Human--Gemini ($\chi^2(1)=266.02$, $p<.001$) and Human--GPT ($\chi^2(1)=203.32$, $p<.001$).


The aggregated voting rates are shown in Fig.~\ref{fig:rq1}. Humans exhibit a smooth, monotonic increase in \textit{n} votes as $\alpha$ increases, reaching near ceiling at $\alpha=1.0$. Gemini follows the same overall trend but saturates below humans, whereas GPT shows a non-monotonic pattern that shifts back toward \textit{so} at $\alpha=1.0$. Overall, VLM response curves (and implied decision boundaries) differ from humans even in the shape-only task. Notably, both VLMs fail to reach ceiling \textit{n} voting at $\alpha=1.0$, suggesting a residual bias toward \textit{so} responses even when the stimulus is visually closest to the \emph{other} character, \textit{n}.

\subsection{Shape-in-Context Word Recognition (RQ2)}
\subsubsection{Sole-Occurrence Context}
\begin{figure*}[t]
    \centering
    \includegraphics[width=\linewidth]{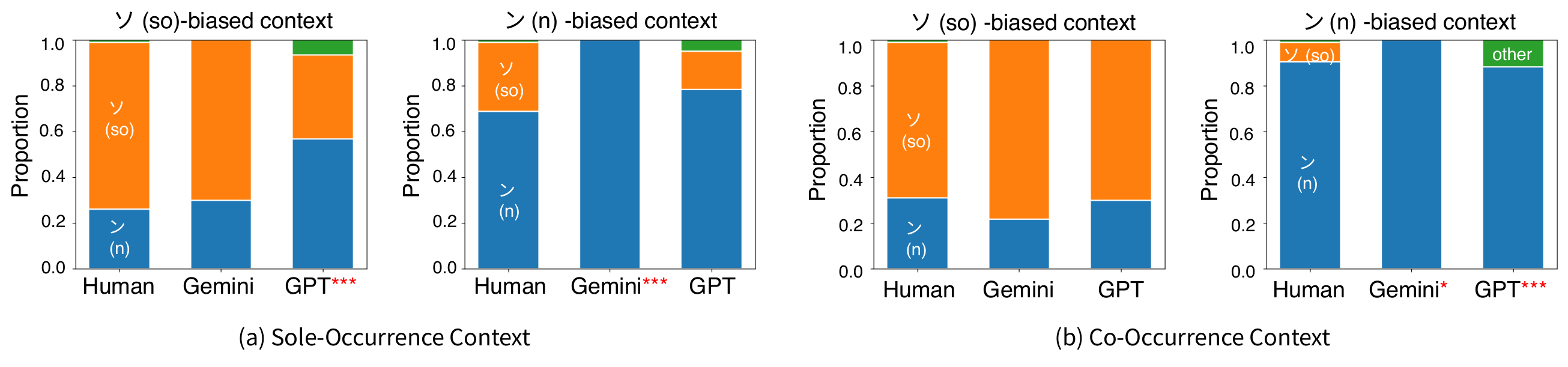}\\[-3mm]
    \caption{Shape-in-context word recognition (RQ2). The target character was replaced with an ambiguous glyph \textit{X} (an interpolation between \textit{so} and \textit{n}) in either so-biased or n-biased word contexts, under (a) sole-occurrence and (b) co-occurrence conditions. Asterisks denote significant differences from humans after Bonferroni correction (\textcolor{red}{*} $p_{\mathrm{adj}}<.05$, \textcolor{red}{***} $p_{\mathrm{adj}}<.001$).}

    \label{fig:rq2}
\end{figure*}

We compared humans and VLMs on response distributions across three categories (\textit{so}, \textit{n}, and other) using Fisher's exact tests on $2\times3$ contingency tables. 
Tests were conducted separately for \textit{so}-biased contexts (the original word has \textit{so} at the target position, i.e., \textit{X} replaces \textit{so}) vs.\ \textit{n}-biased contexts (the original word has \textit{n} at the target position, i.e., $X$ replaces \textit{n}), with Bonferroni correction within each VLM pair ($m=2$).
In \textit{so}-biased contexts, Gemini did not differ from humans ($p_{\mathrm{adj}}=1.000$), whereas GPT did ($p_{\mathrm{adj}}<.001$). 
In \textit{n}-biased contexts, Gemini differed from humans ($p_{\mathrm{adj}}<.001$), whereas GPT did not ($p_{\mathrm{adj}}=.056$).

In \textit{so}-biased contexts of Fig.~\ref{fig:rq2}(a), humans and Gemini mainly chose \textit{so}-consistent readings, whereas GPT leaned slightly toward \textit{n} and differed from humans.
In \textit{n}-biased contexts, humans and GPT mostly chose \textit{n}-consistent readings with occasional \textit{so} responses, while Gemini selected \textit{n} almost exclusively, consistent with a strong \textit{n}-ward tendency in this setting. 
Notably, both VLMs shifted toward \textit{n} in \textit{n}-biased contexts despite their different shape-only tendencies in RQ1 (GPT favored \textit{so}, while Gemini was closer to chance), suggesting that word-level presentation can substantially influence VLM judgments beyond shape-only cues.

\subsubsection{Co-Occurrence Context}
We used the same analysis as in the sole-occurrence condition to compare humans and VLMs (Fisher's exact tests with Bonferroni correction).
In \textit{so}-biased contexts, neither Gemini nor GPT differed from humans ($p_{\mathrm{adj}}=.517$ and $p_{\mathrm{adj}}=1.000$). 
In \textit{n}-biased contexts, both Gemini and GPT differed from humans ($p_{\mathrm{adj}}=.048$ and $p_{\mathrm{adj}}<.001$).


Figure~\ref{fig:rq2}(b) shows that co-occurring unambiguous characters substantially change responses to the ambiguous \textit{X}.
In \textit{so}-biased contexts, humans and both VLMs produced similar distributions with no significant differences, suggesting that within-word character cues support \textit{so}-consistent readings; notably, GPT becomes more human-aligned than in sole-occurrence. 
In \textit{n}-biased contexts, humans mostly selected \textit{n} with a small fraction of \textit{so} responses, whereas Gemini responded almost exclusively with \textit{n}. GPT differed from humans overall, but its proportion of \textit{n} responses is close to humans. 
Overall, adding within-word character cues tends to move VLM behavior closer to human judgments, although model-specific tendencies can persist.



\section{Discussion}
Across our two experiments, we observe qualitative human--VLM alignment that is not captured by accuracy alone. For RQ1, the shape-only task shows that VLM response patterns can differ from humans even in visually simple two-class discrimination: model response curves often transition more gradually than human choices and do not consistently reach ceiling at visually unambiguous endpoints. 
These findings reflect differences in decision behavior under controlled ambiguity, rather than implying that human judgments serve as a normative standard.
For RQ2, embedding the same ambiguous glyphs in words changes behavior, but VLM responses are not uniformly human-aligned: alignment improves in some conditions (most clearly when co-occurrence cues are strong), yet model-specific tendencies persist. Together, these findings suggest that minimal-context inputs are useful as a diagnostic for human--VLM alignment, because providing context can change VLM behavior.
Therefore, to meaningfully assess human--VLM alignment, it is important to observe models both under deliberately minimal-context conditions and under contextualized conditions, rather than relying on either alone.

A key next step is to identify what contextual information drives these shifts in VLM behavior. In particular, to disentangle word-meaning effects from simple co-occurrence cues, we can test pseudo-words (or character-shuffled strings) that preserve versus remove co-occurrence structure within the same word. This would clarify when VLM judgments are driven by word meaning versus local visual evidence, and under which controlled context conditions human--VLM alignment improves or breaks down.

More broadly, this work suggests a gap between strong performance on evaluations with rich context and behavior under deliberately minimal-context inputs. Recent benchmarks highlight growing capability on expert-level problems~\cite{yue2024mmmu,rein2024gpqa}, while other work reports failures on visually straightforward or fine-grained judgments that are easy for humans~\cite{huynh2025vision,deng2025words}. Our results add that accuracy alone can miss important differences in how models resolve perceptual ambiguity, motivating pairing standard benchmark-style evaluations with simple, controlled tests that vary context when discussing human--VLM alignment.

\section{Conclusion and Future Work}
We addressed two research questions about human--VLM alignment under controlled visual ambiguity. For \textbf{RQ1}, we found that VLM shape-based decision boundaries in the shape-only task do not match human boundaries: model response curves transitioned more gradually and did not reliably saturate at visually unambiguous endpoints. For \textbf{RQ2}, we found that embedding the same ambiguous character in words can shift VLM judgments toward human choices, but alignment depends on the available cue and the model: strong within-word co-occurrence cues tended to improve alignment, yet model-specific tendencies persisted across context conditions.

In conclusion, our results show that minimal-context inputs provide a useful diagnostic for human--VLM alignment, and that adding context can shift VLM behavior. Therefore, meaningful assessment of human--VLM alignment should examine both deliberately minimal-context and contextualized conditions, rather than relying on either alone.

\begin{acks}
Generative AI tools (ChatGPT, Gemini) were used for editing and proofreading the text. The authors reviewed and take full responsibility for the content.
\end{acks}


\bibliographystyle{ACM-Reference-Format}
\bibliography{ref}

@String{Computing = "Computing" }

@String{Computer = "{IEEE} Computer" }

@String{Springer = "Springer-Verlag" }

@inproceedings{
higgins2017betavae,
title={beta-{VAE}: Learning Basic Visual Concepts with a Constrained Variational Framework},
author={Irina Higgins and Loic Matthey and Arka Pal and Christopher Burgess and Xavier Glorot and Matthew Botvinick and Shakir Mohamed and Alexander Lerchner},
booktitle={International Conference on Learning Representations (ICLR 2017)},
year={2017},
}

@article{liu2024ocrbench,
  title={{OCRBench}: on the hidden mystery of ocr in large multimodal models},
  author={Liu, Yuliang and Li, Zhang and Huang, Mingxin and Yang, Biao and Yu, Wenwen and Li, Chunyuan and Yin, Xu-Cheng and Liu, Cheng-Lin and Jin, Lianwen and Bai, Xiang},
  journal={Science China Information Sciences},
  volume={67},
  number={12},
  pages={220102},
  year={2024},
  publisher={Springer}
}

@article{fu2024ocrbench,
  title={{OCRBench} v2: An improved benchmark for evaluating large multimodal models on visual text localization and reasoning},
  author={Fu, Ling and Kuang, Zhebin and Song, Jiajun and Huang, Mingxin and Yang, Biao and Li, Yuzhe and Zhu, Linghao and Luo, Qidi and Wang, Xinyu and Lu, Hao and others},
  journal={arXiv preprint arXiv:2501.00321},
  year={2024}
}

@article{firestone2020performance,
  title={Performance vs. competence in human--machine comparisons},
  author={Firestone, Chaz},
  journal={Proceedings of the National Academy of Sciences},
  volume={117},
  number={43},
  pages={26562--26571},
  year={2020},
  publisher={National Academy of Sciences}
}

@inproceedings{palmen2023bold,
  title={How bold can we be? The impact of adjusting font grade on readability in light and dark polarities},
  author={Palm{\'e}n, Hilary and Gilbert, Michael and Crossland, David},
  booktitle={Proceedings of the 2023 CHI Conference on Human Factors in Computing Systems (CHI 2023)},
  pages={1--11},
  year={2023}
}

@article{yasuhara1978category,
  title={Category boundary effect for grapheme perception},
  author={Yasuhara, Makoto and Kuklinski, Theodore T},
  journal={Perception \& Psychophysics},
  volume={23},
  number={2},
  pages={97--104},
  year={1978},
  publisher={Springer}
}

@article{mcclelland1981interactive,
  title={An interactive activation model of context effects in letter perception: I. An account of basic findings.},
  author={McClelland, James L and Rumelhart, David E},
  journal={Psychological review},
  volume={88},
  number={5},
  pages={375},
  year={1981},
  publisher={American Psychological Association}
}

@article{wallace2022towards,
  title={Towards individuated reading experiences: Different fonts increase reading speed for different individuals},
  author={Wallace, Shaun and Bylinskii, Zoya and Dobres, Jonathan and Kerr, Bernard and Berlow, Sam and Treitman, Rick and Kumawat, Nirmal and Arpin, Kathleen and Miller, Dave B and Huang, Jeff and others},
  journal={ACM Transactions on Computer-Human Interaction (TOCHI)},
  volume={29},
  number={4},
  pages={1--56},
  year={2022},
  publisher={ACM New York, NY}
}

@inproceedings{cai2024cor,
  title={COR Themes for Readability from Iterative Feedback},
  author={Cai, Tianyuan and Niklaus, Aleena Gertrudes and Kerr, Bernard and Kraley, Michael and Bylinskii, Zoya},
  booktitle={Proceedings of the 2024 CHI Conference on Human Factors in Computing Systems (CHI 2024)},
  pages={1--23},
  year={2024}
}

@inproceedings{li2019impact,
  title={The impact of web browser reader views on reading speed and user experience},
  author={Li, Qisheng and Morris, Meredith Ringel and Fourney, Adam and Larson, Kevin and Reinecke, Katharina},
  booktitle={Proceedings of the 2019 CHI Conference on Human Factors in Computing Systems (CHI 2019)},
  pages={1--12},
  year={2019}
}

@inproceedings{niklaus2023digital,
  title={Digital reading rulers: Evaluating inclusively designed rulers for readers with dyslexia and without},
  author={Niklaus, Aleena Gertrudes and Cai, Tianyuan and Bylinskii, Zoya and Wallace, Shaun},
  booktitle={Proceedings of the 2023 CHI Conference on Human Factors in Computing Systems (CHI 2023)},
  pages={1--17},
  year={2023}
}

@inproceedings{rzayev2021reading,
  title={Reading in VR: The effect of text presentation type and location},
  author={Rzayev, Rufat and Ugnivenko, Polina and Graf, Sarah and Schwind, Valentin and Henze, Niels},
  booktitle={Proceedings of the 2021 CHI Conference on Human Factors in Computing Systems (CHI 2021)},
  pages={1--10},
  year={2021}
}

@inproceedings{de2024caption,
  title={Caption Royale: Exploring the design space of affective captions from the perspective of deaf and hard-of-hearing individuals},
  author={de Lacerda Pataca, Calu{\~a} and Hassan, Saad and Tinker, Nathan and Peiris, Roshan Lalintha and Huenerfauth, Matt},
  booktitle={Proceedings of the 2024 CHI Conference on Human Factors in Computing Systems (CHI 2024)},
  pages={1--17},
  year={2024}
}

@article{singh2025openai,
  title={OpenAI GPT-5 System Card},
  author={Singh, Aaditya and Fry, Adam and Perelman, Adam and Tart, Adam and Ganesh, Adi and El-Kishky, Ahmed and McLaughlin, Aidan and Low, Aiden and Ostrow, AJ and Ananthram, Akhila and others},
  journal={arXiv preprint arXiv:2601.03267},
  year={2025}
}

@article{team2023gemini,
  title={Gemini: a family of highly capable multimodal models},
  author={Team, Gemini and Anil, Rohan and Borgeaud, Sebastian and Alayrac, Jean-Baptiste and Yu, Jiahui and Soricut, Radu and Schalkwyk, Johan and Dai, Andrew M and Hauth, Anja and Millican, Katie and others},
  journal={arXiv preprint arXiv:2312.11805},
  year={2023}
}

@article{cutler2024word,
  title={The word superiority effect overcomes crowding},
  author={Cutler, June and Bodet, Alexandre and Rivest, Jos{\'e}e and Cavanagh, Patrick},
  journal={Vision Research},
  volume={222},
  pages={108436},
  year={2024},
  publisher={Elsevier}
}

@article{reicher1969perceptual,
  title={Perceptual recognition as a function of meaningfulness of stimulus material.},
  author={Reicher, Gerald M},
  journal={Journal of experimental psychology},
  volume={81},
  number={2},
  pages={275},
  year={1969},
  publisher={American Psychological Association}
}

@article{wheeler1970processes,
  title={Processes in word recognition},
  author={Wheeler, Daniel D},
  journal={Cognitive Psychology},
  volume={1},
  number={1},
  pages={59--85},
  year={1970},
  publisher={Elsevier}
}

@inproceedings{yue2024mmmu,
  title={Mmmu: A massive multi-discipline multimodal understanding and reasoning benchmark for expert agi},
  author={Yue, Xiang and Ni, Yuansheng and Zhang, Kai and Zheng, Tianyu and Liu, Ruoqi and Zhang, Ge and Stevens, Samuel and Jiang, Dongfu and Ren, Weiming and Sun, Yuxuan and others},
  booktitle={Proceedings of the IEEE/CVF Conference on Computer Vision and Pattern Recognition (CVPR 2024)},
  pages={9556--9567},
  year={2024}
}

@inproceedings{rein2024gpqa,
  title={Gpqa: A graduate-level google-proof q\&a benchmark},
  author={Rein, David and Hou, Betty Li and Stickland, Asa Cooper and Petty, Jackson and Pang, Richard Yuanzhe and Dirani, Julien and Michael, Julian and Bowman, Samuel R},
  booktitle={First Conference on Language Modeling},
  year={2024}
}

@inproceedings{huynh2025vision,
  title={Vision-Language Models Can't See the Obvious},
  author={Huynh, Ngoc Dung and Le-Khac, Phuc H and Para, Wamiq Reyaz and Singh, Ankit and Narayan, Sanath},
  booktitle={Proceedings of the IEEE/CVF International Conference on Computer Vision (CVPR 2025)},
  pages={24159--24169},
  year={2025}
}

@inproceedings{deng2025words,
  title={Words or Vision: Do Vision-Language Models Have Blind Faith in Text?},
  author={Deng, Ailin and Cao, Tri and Chen, Zhirui and Hooi, Bryan},
  booktitle={Proceedings of the Computer Vision and Pattern Recognition Conference (CVPR 2025)},
  pages={3867--3876},
  year={2025}
}

\end{document}